\documentstyle[12pt]{article}
\begin{document}
\baselineskip .3in
\pagestyle{empty}
\sloppy
\newpage
\pagestyle{plain}

\title {Phase Distribution in a Disordered Chain and the Emergence
of a Two-parameter Scaling in the Quasi-ballistic to the Mildly
Localized Regime}

\author{Asok K. Sen$^*$ \\
{\it LTP Section, Saha Institute of Nuclear Physics}\\
{\it 1/AF Bidhannagar, Calcutta 700 064, India}}

\date{\today}
\maketitle

\begin{abstract}
We study the phase distribution of the complex reflection coefficient
in different configurations as a disordered 1D system evolves in length,
and its effect on the distribution of the 4-probe resistance $R_4$.
The stationary ($L \rightarrow \infty$) phase distribution is almost
always strongly non-uniform and is in general double-peaked with their
separation decaying algebraically with growing disorder strength to
finally give rise to a single narrow peak at infinitely strong disorder.
Further in the length regime where the phase distribution still evolves
with length (i.e., in the {\it quasi-ballistic} to the {\it mildly
localized} regime), the phase distribution affects the distribution of
the resistance in such a way as to make the mean and the variance of
$log(1+R_4)$ diverge independently with length with different exponents.
As $L \rightarrow \infty$, these two exponents become identical (unity).
Obviously, these facts imply two relevant
parameters for scaling in the quasi-ballistic to the mildly localized
regime finally crossing over to one-parameter scaling in the strongly
localized regime.

\end{abstract}

{PACS numbers: 72.10.Fk, 72.15.Rn, 73.20.Fz, 73.20.-b}

$^*${\it e-mail address: asok@hp2.saha.ernet.in}

\newpage

Evolution of the reflectance (or, the two-probe resistance, $R$) and
the phase ($\phi$)
of the complex reflection coefficient [$r=R^{1/2} exp(i\phi)$] with
length ($L$) due to backscattering in a disordered quantum conductor
are intimately connected.  Indeed one is coupled to the other as
demonstrated clearly, e.g., in the coupled differential equations
for these quantities obtained by using the invariant imbedding method
\cite{hein1, rd, bw}.  It may be noted here that for a single disordered
sample, as opposed to an ensemble of them, there is no finite phase
decoherence length at zero temperature.  Since phase coherence is of
paramount importance in disordered systems, the important object to
look for is the distribution of the phases in different configurations
and its evolution with length ($L$).  It is thus intriguing to note that
even though a lot of work has been done for the last three to four
decades on the resistance and its moments (and on its probability
density as well), comparatively little has been
done on the phase distribution and its evolution with length ($L$).
While rigorous results exist for exponential localization in 1D, i.e.,

$$ R(L) \sim exp(2 \alpha L), \eqno (1)$$

\noindent in the $L \rightarrow \infty$ limit ($\xi=1/\alpha$ being the
localization
length), and field-theoretic and numerical methods indicate an insulator
to metal transition only above 2D (at least in the absence of any magnetic
or other fields which break the time-reversal invariance and of spin-orbit
scattering \cite{fast}), no analytic
or otherwise fully definitive result for the phase distribution covering
the complete domain of real space or disorder strength exists even in 1D.
Notwithstanding this fact, almost all analytic works assume random 
\cite{ataf} (uniform) distribution of the phase angle between 0 and $2\pi$
to solve for the distribution of the resistance.  It is sometimes called the
random phase model.  In the first half of this Letter, we first concentrate
on the evolution of the phase distribution (in a disordered 1D system) with
$L$ and study how the peaks in the stationary ($L \rightarrow \infty$)
phase distribution behave as the disorder strengths are varied from
weak to infinitely strong.  Since the phase distribution for a very large
length keeps changing with disorder strength, clearly  a large length scale
should not necessarily imply  a large disorder scale even in 1D.  In the
latter half of this work, we come back to moderate length scales where the
phase distribution is still evolving with length and study its effect on
the first two moments of the logarithmic resistance.  The main result of
this work is that in this regime, the above two moments diverge with
different exponents contrary to the case of a $log$-$normal$ distribution
where there should have been a single exponent.  Also, interestingly,
for very large length scales ($L>>\xi$), the above two moments diverge
with the same exponent (as in the case of a $log$-$normal$ distribution)
even though the stationary phase distribution is far from uniform.  Hence
our conclusion below that as long as the non-uniform phase distribution
does not settle down to its stationary limit, the logarithmic resistance
distribution gets affected in such a way that its description requires
at least two independent scaling parameters and not one.

\smallskip

The first definitive work on the issue of the phase distribution seems to
be the numerical transfer matrix work of Stone, Allan and Joannopoulos
\cite{saj} (referred to as SAJ from now on) which indicates very clearly
that this distribution \cite{f1} is far from uniform except in the {\it
diffusive}
(disordered metallic) regime.  As a matter of fact, it was shown by SAJ
that for disorder or length scale much smaller than that for the diffusive
regime (i.e., {\it quasi-ballistic} in the current terminology), the phase
angle distribution is doubly peaked around $\phi={\pi \over 2}$ and
$\phi={3\pi \over 2}$ respectively.  We note that very recently this 
result was obtained analytically by Heinrichs \cite{hein2} by solving
the invariant imbedding equations exactly in the limit $L<<(2k_F)^{-1}
<\xi$, where $k_F$ is the Fermi wave-vector of the incoming electrons
from a semi-infinite perfect lead attached to the source reservoir
of the electrons.  It is only in the regime $(2k_F)^{-1}<<L<\xi$ that
the system is diffusive (or, quasi-metallic) and the
phase distribution is nearly uniform.  This is the so-called {\it
weak localization regime}.  In 1D, the two-probe conductance
(transmittance) in this regime also follows a nearly uniform distribution
\cite{kjs} and thus gives rise to a universal conductance fluctuation
(UCF) appropriate for 1D \cite{gs1}.
But what happens to the phase for $L>>\xi$ and $W \rightarrow \infty$
seems to be somewhat controversial (here $W$ is the strength of the
disorder as measured by the width of the randomness characterizing
the problem).  This is important
because, as we discussed before, two very different distributions in
$\phi$ should affect the stationary ($L \rightarrow \infty$) distribution
of $R$ differently and the effect should be clearly seen in the ensemble-
averaged value of the resistance.  The old SAJ result indicates that the
stationary phase distribution in the strong localization (very large
disorder) limit becomes sharply peaked (single peak) at an angle dependent
only on the Fermi energy of the electron but independent of $W$.  But,
there also exists an opposite view \cite{jay} that in the strong disorder
limit the stationary distribution becomes doubly peaked.

\smallskip

In the first part of this Letter we investigate this controversial issue
in the very large length scale or the fully localized regime (in 1D).
We work within a single band tight binding hamiltonian on a lattice with
a lattice constant $a=1$ to set the length scale, the nearest neighbour
hopping term $V=1$ to set the energy scale, and an uniform random
distribution in the site energy: $P(\epsilon_n) ={1\over W}$, in the
domain $[-{W \over 2},{W\over 2}]$, and zero elsewhere. Thus $W\over V$
serves as a measure of the strength of disorder. 
A transfer matrix method described elsewhere \cite{gs1} has been used to
solve for the complex reflection coefficient $r$ of the sample attached to
semi-infinite perfect leads. The phase $\phi$ is calculated from the real
and imaginary parts of $r$ and properly shifted into the domain $[0,2\pi]$.
For our calculation of phases, we have worked with $5,000-25,000$
configurations (larger the disordered chain length, larger the number),
and chose lengths such that the distribution does not change significantly
on increasing the length further. Thus the histograms we present below do
indeed refer to the stationary distributions of the phase, $P(\phi)$, in
the medium to very strong localization regime.  Further, all the histograms
have been normalized for the purposes of comparison. 

In Fig.1 we show the stationary distribution of phase at the Fermi energy
$E_F=1.6V$ for different disorders ranging from ${W\over V}=3$ to 
${W\over V}=200$.  The localization
length is found to vary from $\xi \simeq 3a$ to $\xi \simeq 0.12a$ in this
range.  There are several features to be noticed from these histograms.
The phase distribution is already very non-uniform for ${W\over V}=3$, and
it has two broad peaks one about twice in altitude than the other.  As the
disorder grows to ${W\over V}=8$ (Fig. 1b), the non-uniformity in $P(\phi)$
becomes more pronounced.  The peaks become narrower and stronger (at the
cost of a reduction in the background).  Individually the weaker peak
grows more in strength, and thus the strength of the two peaks tend to
approach each other.  Most importantly, they come closer as $W$ increases.
This trend continues upto ${W\over V}=60$ (Fig. 1e), where one can still 
distinguish between these two peaks very close in strength and position
in the phase domain.  Finally for ${W\over V}=200$ (Fig. 1f), we cannot 
distinguish between the two peaks for our chosen bin size of
${\Delta \phi \over 2\pi}=0.01$,
and the distribution becomes a very strongly peaked function.  For our
accuracy, this disorder seems to be {\it infinitely} strong.  We did also
look at the position of this single peak for infinitely strong disorder,
and find that it appears at $\phi_\infty=2cos^{-1}{E_F\over 2}$.  For
example, the peak in Fig. 1f appears at $\phi_\infty=$1.29 radian. 
This result may be analytically calculated exactly from the fact that
for an infinitely strong
disorder ($W \rightarrow \infty$), the electron gets almost completely
backscattered from a single (the first in the chain) impurity itself.
One may thus look at a single, infinitely strong disorder in the middle
of an otherwise ordered chain, and obtain the above result exactly.
Interestingly $\phi_\infty$ lies in between the two peaks for any particular
disorder in Figs. 1a-1e.  It may be noted here that the appearence of the
two peaks for large lengths and for intermediate disorder was previously
reported in the literature \cite{mm} as a passing remark.  In that work, 
done only at one $E_F$, the second peak was just a very mild hump, whose
strength was reported to be decreasing (finally vanishingly small) with
disorder and hence all the emphasis was given to the single strong peak
and the variation of its position with $W$.  In contrast, we find that
at any fixed $E_F$, the strength of the weaker peak grows stronger and
the two peaks move (as if attracted) towards each other with increasing
$W$ (see Fig. 1), and finally the two peaks coalesce to give a single
peak for infinitely strong disorders.

\smallskip

Obviously the next question would be how does the two peaks in the
stationary distribution approach each other as $W$ increases.  For this
purpose, we have studied the histograms as in Fig.1 at other energies
and find that as one approaches
closer to the band center (i.e., $|E_F| \rightarrow 0$), the onset
disorder for the appearence of two peaks (in the medium strong disorder
regime) becomes larger.  For example for ${E_F\over V}=1.6$, this onset
value is ${W\over V} \simeq 3$ but for ${E_F\over V}=0.1$, the onset
value is ${W\over V} \simeq 5$.  Further whereas the peaks have unequal
strengths when the Fermi 
energy is away from the band center, on approaching the band center
the two peaks assume near idential strengths even near the onset
disorder.  Thus except for the `mass' center ($\phi_\infty$) and the
relative `masses' of the two peaks, the sequence of histograms at any
energy look quite similar to those in Fig. 1.  For the purposes of
illustration we have chosen both the energies mentioned above.  In
Fig. 2a, we have plotted the relative positions of the left and the right
peaks with respect to $\phi_\infty$ (to look for asymmetry in their
positions) as a function of ${W\over V}$ at the energy ${E_F\over V}=0.1$
and in Fig. 2b we have done the same for ${E_F\over V}=1.6$.  The
excellent linear fits for both the sets and their near-parallelity on
the log-log plot clearly indicate a power law with the same exponent.
The results may be summarized as follows.  For $L >> \xi$, we find that
the asymptotically stationary phase distribution is in general a
two-peaked function with the left and the right peaks at

$$ \phi_{r,l} = \phi_\infty \pm {b_{r,l} \over W^\mu}, \eqno (2) $$

\noindent where $b_l$ and $b_r$ are constants independent of ${W\over V}$
but dependent only on ${E_F\over V}$, whereas $\mu$ is a constant
independent of ${W\over V}$ or ${E_F\over V}$ ($+$ for $r$ and $-$ for
$l$-peaks).  For the
case in Fig. 2a where ${E_F\over V}=0.1$, we have $b_l \simeq b_r=0.50$,
and $\mu=0.85$.  For Fig. 2b where ${E_F\over V}=1.6$, we still have 
$\mu=0.85$, but $b_l=0.37$ and $b_r=0.28$.  Very similar results
are obtained at any other energy.  The inequality of $b_l$ and
$b_r$ away from the band center indicate that the peaks are
asymmetrically separated from their fixed asymptotic ($W \rightarrow
\infty$) position.  This asymmetry seems to be correlated to the
inequality in peak-strengths at finite strengths at finite $W$'s
and for energies away from the band center.  It may be noted that
the eq.(2) indicates that as $W$ increases, the separation 
between the peaks decays as:

$$ \Delta \phi = {b \over W^\mu},    \eqno (3)  $$

\noindent where $b=b_l+b_r$.  Let us now look at a simple consequence.
The renormalization group flow in the one-parameter scaling \cite{aalr}
in 1D would have us believe that increasing the length scale is equivalent
to increasing the disorder scale.  But Eq.(2) implies that different
disorder strengths give rise to different stationary ($L \rightarrow
\infty$) phase distributions and hence indicates that, contrary to the
one-parameter scaling, a large length scale is not completely 
eqivalent to a large disorder scale.

\smallskip

To be more sure of the effects of the still-evolving strongly non-uniform
phase distribution, we study in this latter half of our work, the
evolution of the logarithmic resistance in the domain where stationarity
is not yet reached.  In particular we are interested in testing the
validity or the lack thereof of the one-parameter scaling theory
\cite{aalr}.  Let us explain our motivation clearly.  It is well-known
that resistance (or, conductance) of a disordered system is not
self-averaging in the sense that its distribution has a log-normal
tail even in the metallic regime, and becomes completely log-normal
in the localized regime \cite{akl}.  Here the word log-normal means
that the random variable $u(L)=ln(1+R_4)=-ln T(L)$, where $R_4(L)=
R(L)/T(L)$ is the four-probe resistance and $T(L)=1-R(L)$ is the
two-probe conductance, has a normal distribution.  Now in most of the
previous analytical works \cite{ataf}, an uniform random distribution
for $\phi$ is assumed, and one then obtains the result that the two
parameters of $P_L(u)$, namely the mean and the variance diverge
linearly with $L$.  As in any normal distribution, they diverge with
the same exponent (i.e., exactly unity), and hence one obtains
only one relevant parameter in the scaling
of the resistance.  The main result of this Letter as shown below
is that in the {\it regime where $P_L(\phi)$ is both non-uniform and
is changing with $L$}, the mean and the variance of $u$ diverge with
different exponents and hence in that regime {\it the one-parameter
scaling does not seem to hold}.  Of course this state of affairs does
not hold for $L >> \xi$, and we show that in this regime where
$P_L(\phi)$ is independent of length, i.e., {\it stationary} (although
far from uniform), the mean and the variance of $u(L)$ do
again diverge as $L$ consistent with one-parameter scaling in the very
large length scale or very strong disorder regime.

\smallskip

For our purpose we calculate $R_4(L)$ for 20,000-40,000 configurations
as a function of $L$ at the Fermi energy ${E_F\over V}=1.6$ and a
disorder strength ${W\over V}=1$.  Then we calculate the mean
$<u>$ and the variance $var(u)=<u^2>-<u>^2$ as a function of $L$.
It may be pointed out that in this case $P_L(\phi)$ does not
assume its stationary form, i.e., $P(\phi)$, until a length of
about $L=300$.  From the exponential drop of the average reflectance
we find that the localization length is $\xi \simeq 40$ and the
$P_L(\phi)$ is double peaked (figure not shown here for brevity) even
in the quasi-ballistic as well as the weakly localized regime.
As a result the probability density of
the logarithmic resistance, $P_L(u)$, is strongly asymmetric
in the entire range of length ($L$= 10 - 700) studied even though
the asymmetry gradually becomes smaller with increasing $L$.  Thus
in contrast to all the previous works known to the author, $P_L(u)$
is non-Gaussian even in the weakly localized regime.
More details on this will be discussed elsewhere.  For the purpose
of this Letter, we show in Fig.3 the log-log plot for the $<u>$ and the
$var(u)$ as a function of $L$.  We find that whereas $<u(L)>=0.058L$
indicating pure exponential growth in the entire domain, $var(u)$ first
goes as 0.0075$L^{1.57}$ upto $L \simeq 80$ (2$\xi$) and crosses
over towards $0.11L$ for $L \ge 300$.  It is interesting to note that
the graphs for $<u>$ and $var(u)$ cross each other at $L=\xi \simeq
40$.  Thus $<u>$ and $var(u)$ diverge with two independent exponents,
namely 1 and 1.57 respectively, from the quasi-ballistic through the
quasi-diffusive to the mildly localized regime (upto about $2\xi$).
In this regime, the first two moments of $P_L(u)$, i.e., $<u>$ and
$var(u)$, behave as two independent parameters (as opposed to a
gaussian behaviour) or relevant variables for scaling and hence the
one-parameter scaling does not seem to adequately describe the
transport properties.  Whether these two are the the only relevant
variables in this regime is related to the nature of the deviation of
$P_L(u)$ from gaussianity and needs to be studied more carefully.  Also
it may be noted (Fig. 3) that for $L \ge 7\xi$, the phase distribution
$P(\phi)$ is stationary, and both $<u>$ and $var(u)$ diverge 
identically (i.e., linearly) indicating that there is just
one relevant variable (e.g., only the $<u>$), and the one-parameter
scaling theory is valid for very large length scales.  In this
connection it may be noted that in one of our previous works
\cite{gs2} on conductance and its fluctuations from the quasi-ballistic
to the diffusive regime, we did find the existence of two
independent length scales, namely $\xi$ and a quasi-ballistic length
scale $\Delta(W)$ which is essentially a periodicity length of
conductance oscillations.  It is interesting to note that the effect
of this extra length scale $\Delta$ may also be discerned only upto
$L \simeq \xi$, since the conductance oscillations become imperceptible
beyond that length.  The two independent parameters in the
form of length scales should show up in conductance measurements.
The measurements could be quite tricky though since the amplitude
of oscillation depends upon how much the hopping term (or, the
effective electronic mass) in the sample differs from that in the
connecting leads.  If this difference is not large enough, the
UCF may mask the oscillations, i.e., the other length scale
completely \cite{gs2}. 

\smallskip

We stress again that our aim was not to find the inadequacy of the
one-parameter scaling in the thermodynamic limit, but rather to find
why and where could this breakdown occur.  Our findings suggest that this
occurs for moderate disorders and near the onset of the metal-insulator 
transition where the phase distribution still evolves with length.  It is
true that a quantum phase transition from an insulator to a metal does not
take place in 1D in the thermodynamic limit, and hence the effect mentioned
above looks like a crossover effect.  But it is also true that on
renormalization from a small to a large length scale, the 1D system flows
from the
weakly localized to the strongly localized regime qualitatively similarly
to that for a 2D or for a $(2+\epsilon)$-dimensional system on the localized
side of the mobility edge $E_c$ (but initially very close to it).  On
qualitative grounds, we believe that the phase distribution will evolve
in the latter two cases just as in 1D.  Thus we expect a two-parameter
scaling similar to the one discussed above in 2D as well as in $(2+\epsilon)$
dimension on the localized side.  The non-trivial second exponent
(of $var(u)$) would be Fermi-energy-dependent as in above.  Further
we conjecture that on the extended side of $E_c$ but close to it, $<u>$ and
$var(u)$ would diverge independently and one would find two independent
exponents in the distribution.  But on renormalization, one would again
scale far away from $E_c$ on the extended side (since one is not
sitting exactly on the $E_c$), the behavior would again {\it crossover} to
a qualitatively different one, namely that $<u>$ would diverge
logarithmically and $var(u)$ would tend to vanish (consistent with an Ohmic
behavior, which is our guiding principle in this asymptotic limit).
Further work needs to be done in this area.

As a further support, we indicate a possible relation of our results to
a recent novel work \cite{klaa} on 3D disordered systems, where it has
been shown using supersymmetric non-linear sigma model, that the {\it level
spacing distribution close to $E_c$ is quite non-standard} in the sense
that the distribution is non-Wigner-Dyson in the disordered metallic
side and non-Poissonian in the insulating side.  The first thing to note is
that this behavior would also look like a crossover phenomenon in the length
scale (as discussed above) since on renormalization one goes far away from
$E_c$ and the non-standard behavior would predictably crossover to one of
the standard forms (Wigner-Dyson or Poisson).  Second, if one notes that
the electronic conduction (say, using Kubo formula) would require the
hopping of the electron between two energy levels on either side
of the Fermi energy, it becomes clear that this process really samples
the level spacing distribution around $E_F$.  Thus if $E_F$ is close
to $E_c$ and if something unusual is happening in the level spacing
distribution around the latter, the distribution and the scaling
properties of the conductance should also become unusual there (e.g.,
neither Ohmic nor log-normal as in our case), and $vice$-$versa$.  The
relation between the level spacing distribution and the conductance
distribution would be explored further in the future.

\smallskip

The author acknowledges very useful discussions with B.I. Halperin.

\newpage

{\bf Figure Captions:}

{\bf Fig. 1} Normalized histograms representing the stationary
($L \rightarrow \infty$) phase distributions at $E_F/V=1.6$ and
at an increasing sequence of disorder in (a) to (f).

{\bf Fig. 2} Double logaritmic plots of the relative positions of the
left and the right peaks ($|\phi_{l,r}-\phi_\infty|$) in the
stationary ($L \rightarrow \infty$) phase distribution as a function
of the disorder strength $W/V$ at two different energies: (a) for
$E_F/V=0.1$, the peaks are nearly symmetrically placed, and (b) for
$E_F/V=1.6$, the peaks are asymmetrically placed about $\phi_\infty$.

{\bf Fig. 3} Double logarithmic plots of the mean and the variance
of the logarithmic four-probe resistance, $u=log(1+R_4)$ as a
function of length $L$ for $E_F/V=1.6$. While $<u>$ diverges as
$L$ all the way from the quasi-ballistic to the strongly localized
regime, $var(u)$ diverges with an exponent of 1.57 in the regime
from the quasi-ballistic to the mildly localized ($L \simeq 2\xi$)
and then crosses over to a divergence as L in the strongly
localized regime.


\bigskip

\begin{thebibliography}{9}

\bibitem{hein1} N. Kumar, Phys. Rev B{\bf 31}, 5513 (1985); J. Heinrichs,
Phys. Rev. B{\bf 33}, 5261 (1986)

\bibitem{rd} R. Rammal and B. Doucot, J. Phys.(Paris){\bf 83}, 509 (1987)

\bibitem{bw} R. Bellman and G.M. Wing, {\it An introduction to invariant
imbedding} (Wiley, New York, 1976)

\bibitem{fast} U. Fasternath, Physica A {\bf 189}, 27 (1992)

\bibitem{ataf} See for example P.W. Anderson, D.J. Thouless, E. Abrahams,
and D.S. Fisher, Phys. Rev. B{\bf 22}, 3519 (1980)

\bibitem{saj} A.D. Stone, D.C. Allan, and J.D. Joannopoulos, Phys. Rev.
B{\bf 27}, 836 (1983)

\bibitem{f1} Actually the phase defined by SAJ \cite{saj} is not quite
that of the total reflectance, but of a closely related quantity.

\bibitem{hein2} J. Heinrichs, J. Phys.:Condens. Matter {\bf 7}, 6291 (1995)

\bibitem{kjs} A. Kar Gupta, A.M. Jayannavar, and A.K. Sen, J. Phys. I
France {\bf 3}, 1671 (1993)

\bibitem{gs1} S. Gangopadhyay and A.K. Sen, Phys. Rev. B{\bf 46}, 4020
(1992).  Note that the UCF reported here persists upto a length
slightly larger than $\xi$.

\bibitem{jay} A.M. Jayannavar, Pramana-J. Phys. {\bf 36}, 611 (1991)

\bibitem{mm} S.K. Manna and A. Mookerjee, Int. J. Mod. Phys. B {\bf 6},
1517 (1992)

\bibitem{aalr} E. Abrahams, P.W. Anderson, D.C. Licciardello and
T.V. Ramakrishnan, Phys. Rev. Lett. {\bf 42}, 673 (1979)

\bibitem{akl} B.L. Al'tshuler, V.E. Kravtsov, I.V. Lerner, in {\it
Mesoscopic Phenomena in Solids}, eds. B.L. Al'tshuler, P.A. Lee and
R.A. Webb, pp. 449-521 (North Holland, Amsterdam, 1991)

\bibitem{gs2} M.N. Ganguli and A.K. Sen, Phys. Rev. B{\bf 52},
17342 (1995)

\bibitem{klaa} V.E. Kravtsov, I.V. Lerner, B.L. Al'tshuler and
A.G. Aronov, Phys. Rev. Lett. {\bf 72}, 888 (1994)

\end{thebibliography}
\end{document}